# Probing topological insulators surface states via plasma-wave Terahertz detection

Leonardo Viti[1], Dominique Coquillat[2] Antonio Politano[3], Konstantin A. Kokh[4], Ziya S.Aliev[5,6], Mahammad B. Babanly[5], Oleg E. Tereshchenko[7,8], Wojciech Knap,[2,9] Evgueni V. Chulkov[10,11,12] and Miriam S. Vitiello[1]

*[1] NEST, Istituto Nanoscienze – CNR and Scuola Normale Superiore, Piazza San Silvestro 12, Pisa, I-56127*
*[2] Laboratoire Charles Coulomb (L2C), UMR 5221 CNRS-Univ. Montpellier 2, Montpellier, France.*
*[3] Università degli Studi della Calabria, Dipartimento di Fisica, via Ponte Bucci, 87036 Rende (CS), Italy*
*[4] V.S. Sobolev Institute of Geology and Mineralogy, Siberian Branch, Russian Academy of Sciences, Koptyuga pr. 3, Novosibirsk, 630090 Russian Federation*
*[5] Institute of Catalisis and Inorganic Chemistry Department, Azerbaijan National Academy of Sciences, AZ-1143 , Baku, Azerbaijan*
*[6]Institute of Physics, Azerbaijan National Academy of Science, AZ1143 Baku, Azerbaijan*
*[7]Institute of Semiconductor Physics, Siberian Branch, Russian Academy of Sciences, pr. Akademika Lavrent'eva 13, Novosibirsk, 630090 Russian Federation*
*[8]Novosibirsk State University, ul. Pirogova 2, Novosibirsk, 630090 Russian Federation*
*[9]Institute of High Pressure Institute Physics Polish Academy of Sciences Warsaw Poland*
*[10] Donostia International Physics Center (DIPC), CFM-MPC, Centro Mixto CSIC-UPV/EHU, Departamento de Fisica de Materiales, UPV/EHU, E-20080 San Sebastian, Spain*
*[11]Tomsk State University, Pr. Lenin 36, 634050 Tomsk, Russian Federation*
*[12]Saint Petersburg State University, Saint Petersburg 198504, Russian Federation*

**Topological insulators (TIs) represent a novel quantum state of matter, characterized by edge or surface-states, showing up on the topological character of the bulk wave-functions. Allowing electrons to move along their surface, but not through their inside, they emerged as an intriguing material platform for the exploration of exotic physical phenomena, somehow resembling the graphene Dirac-cone physics, as well as for exciting applications in optoelectronics, spintronics, nanoscience, low-power electronics and quantum computing. Investigation of topological surface states (TSS) is conventionally hindered by the fact that in most of experimental conditions the TSS properties are mixed up with those of bulk-states. Here, we devise a novel tool to unveil TSS and to probe related plasmonic effects. By engineering $Bi_2Te_{(3-x)}Se_x$ stoichiometry, and by gating the surface of nanoscale field-effect-transistors, exploiting thin flakes of $Bi_2Te_{2.2}Se_{0.8}$ or $Bi_2Se_3$, we provide the first demonstration of room-temperature Terahertz (THz) detection mediated by over-damped plasma-wave oscillations on the "activated" TSS of a $Bi_2Te_{2.2}Se_{0.8}$ flake. The reported detection performances allow a realistic exploitation of TSS for large-area, fast imaging, promising superb impacts on THz photonics.**

The raising interest in THz radiation (loosely defined as the 0.1−10 THz frequency range, 3000 - 30 μm wavelength range) and in the related application-oriented issues in everyday life, requires the progressive development of sensitive and performing systems exploiting powerful, stable and coherent sources as well as fast, sensitive and portable photodetectors. These key priorities prompted in the last decade a major surge of interdisciplinary research, encompassing the investigation of different technologies in-between optics and microwave electronics, different physical mechanisms and a large variety of material systems offering *ad-hoc* properties to target the expected performance and functionalities.

In this context, a wealth of room-temperature (RT) detection technologies[1] have been recently proposed and implemented, e.g, focal plane arrays (FPA) of microbolometers based on semiconductor or superconductor thermal sensing elements, photoacoustic conversion detectors based on carbon nanotubes[2] and fast nonlinear rectifying electronic components such as Shottky diodes, high-electron-mobility transistors (HEMT) and field effect transistors (FET)[1,3] exploiting III-V semiconductors,[3] semiconductor nanowires,[4] carbon nanotubes,[5] and more recently bi-dimensional (2D) materials, like graphene.[6-9]

FET THz detectors conventionally operate at RT in the overdamped plasma waves regime (or distributed self-mixing regime) in which THz rectification appears when the electromagnetic *ac* field, coupled to the source (S) and gate (G) electrodes, simultaneously modulates the carrier density and drift velocity.[10,11] The resulting current exhibits a continuous (dc) component, whose magnitude is proportional to the intensity of the incoming radiation, and can be measured at the drain (D) contact either in a short circuit (photocurrent mode) or in an open circuit (photovoltage mode) configuration. This mechanism depends on: *i*) the gate capability to modulate the channel density, i.e. the FET's transconductance ($g_m$), and *ii*) the asymmetric feeding of the *ac* field into the channel. The first condition can be achieved by increasing the gate-to-channel capacitance. For the second condition to be fulfilled, the S and D electrodes must have different geometries or the channel itself conceived to be intrinsically asymmetric.[12]

Under this picture, if the electronic properties of a 2D active-channel material are properly engineered, THz radiation may activate over-damped plasma waves, leading to detection that can be a tool to unveil/investigate specific surface states -related plasmonic phenomena.

Topological insulators can be, in this perspective, a very intriguing material platform.[13-16] TIs are semiconductors in the bulk, while they exhibit metallic conduction at the surface, due to the existence of well-defined topological surface states (TSS). TSS form a Dirac cone around the Γ point and their presence is determined only by the bulk electronic structure, in which the strong spin-orbit interaction produces a symmetry inversion in the band gap order, giving rise to a topologically non-trivial band configuration.[13]
The activation and exploitation of TSS in an active device is a very demanding task, although one can easily release the ground-breaking potential for a variety of applications,[14,17,18,19] and appealingly, even for light detection: i) the TSS are gapless, enabling charge carrier generation by light absorption over a very wide energy spectrum - including the ultraviolet, visible, near-infrared, mid-infrared, and THz spectral regimes - with a signal-to-noise ratio (SNR), which, via thickness control, can largely exceed that of conventional photo-detecting material (as bulk $Hg_{1-x}Cd_xTe$)[1]; ii) the optical and electronic properties can be engineered by playing with the material stoichiometry;[20,21] iii) the TSS mobility can reach values even larger than graphene due to the topological protection that prevents backscattering effects;[22] vi) with almost the same high-absorbance of graphene, they can exploit a tunable surface band gap which is a major need in optoelectronics;[9] v) the two-dimensional electron gas (2DEG) arising from the TSS of three-dimensional (3D) TIs supports a collective excitation (Dirac plasmon) in the THz range.[23]

Presently, TSS have been only revealed via angle-resolved photoemission spectroscopy,[20,24] in TI samples showing sufficiently high structural quality (both stoichiometric and crystallographic), i.e., a Fermi level shifted from the bulk band gap and pinned by the bulk states occupied by doping electrons or holes.

Conversely, TSS are not accessible via conventional transport measurements;[25] as an example, the TSS conductivity in samples exploiting $Bi_2Se_3$, a prototypical TI material, is usually overshadowed by the high residual bulk conductivity. Furthermore, the presence of Se vacancies favors the formation of Bi-O bonds,[26] which induces rapid oxidation and surface degradation.

Recently, the optimized use of the Bridman-Stockbarger method in vacuum/inert atmosphere,[27] proved to be a valuable tool to reduce the amount of vacancies in TI samples, allowing chemical inertness toward surface oxidation and the capability of a fine tuning of the Fermi level,[27] opening appealing perspectives for band-structure engineering of TI heterostructures. In this context, bismuth chalcogenides alloys have established themselves as intriguing building blocks for optoelectronic applications at RT: their TSS carries can reach very high-mobility (9000 $cm^2/Vs$),[28] and superb carrier density tunability (~100%) through electrical gating.[29] This makes these materials ideal to investigate THz detection mediated by TSS in the over-damped plasma-wave regime, providing an intriguing way to unveil the TSS, optically.

In this letter, we report on the first demonstration of THz detection in top-gated nanometer FETs exploiting thin $Bi_2Te_{(3-x)}Se_x$ flakes. This material platform is very interesting for TSS investigations because on one side, the Te-free alloy, $Bi_2Se_3$, shows a topologically non-trivial energy gap of 0.3 eV, about ten times higher than thermal excitation at RT,[15] and on the other side, recent experiments[20] unveiled that the Fermi level of high Te-content alloys, like $Bi_2Te_{2.2}Se_{0.8}$, crosses only the TSS Dirac cone, meaning that its TSS behave as an ideal 2D electron gas. Here, we compare THz detection in antenna-coupled FETs having as active channel the surfaces of $Bi_2Se_3$ (sample A) or $Bi_2Te_{2.2}Se_{0.8}$ (sample B).

Single-crystalline ingots of $Bi_2Se_3$ and $Bi_2Te_{2.2}Se_{0.8}$ were grown from melt by the vertical vacuum/inert atmosphere optimized Bridgman-Stockbarger method (see Methods). Exfoliated flakes were then identified via optical and scanning electron (Figs.1a-b) microscopy (SEM) and characterized with atomic force microscopy (AFM) (Figs.1c-d) and micro-Raman spectroscopy (Figs.1e-f) (See Methods). Thin flakes ($h$ ~ 10 nm, corresponding to ~ 7 quintuple layers) were then individually contacted with proper adhesion layer/metal sequences to define the S and D electrodes via aligned electron beam lithography (EBL) (see Methods). The S and G electrodes were then patterned in the shape of a half 100° bow-tie antenna, introducing the required asymmetry for over-damped plasma-wave detection (see Methods). The SEM images of the device are shown in Figs. 2a-d, together with the conceived device schematics (Fig. 2e). Flakes having total length < 2 μm have been selected as active non-linear elements of our FETs, in order to reduce both the parasitic capacitance and the resistances in the un-gated transistor regions.

Figures 3a-b shows the transconductance characteristics of sample A (a) and B (b) measured while keeping fixed the source-drain voltage $V_{SD}$ = 1 mV, and while maintaining the gate bias ($V_G$) below the breakdown voltage of the top-gate oxide (estimated to be ≈ 15 V). The unveiled n-type FET behaviours are likely due to Se vacancies, crystal defects or environmental doping. In a rigorous approach, two types of carriers (TSS and bulk) should be considered. Below, we roughly extracted carrier density ($n$) and active channel material mobility ($\mu$) as $n = C_{gc}V_{th}/(e \cdot h \cdot A_g)$ where $C_{gc}$ is the gate-to-channel capacitance, $V_{th}$ is the FET threshold voltage, $e$ is the electron charge and $A_g$ is the area of the gated region and $\mu = (ne\rho)^{-1} = L_g^2/(C_{gc}V_{th}R_g)$ where $\rho$ is the channel resistivity at $V_G = 0$, $R_g$ is the resistance of the gated region and $L_g$ the gate length (see Methods). We obtained $n = 4.7 \times 10^{18}$ cm$^{-3}$ and $\mu = 136$ cm$^2$/Vs for $Bi_2Se_3$ (sample A) and $n = 2.8 \times 10^{19}$ cm$^{-3}$ and $\mu = 210$ cm$^2$/Vs for $Bi_2Te_{2.2}Se_{0.8}$ (sample B). Sample A shows an almost one order of magnitude higher conductivity modulation with $V_G$, with an inaccessible charge neutrality point, as expected for oxide-gated FETs at RT.[30] In both cases, the huge hysteresis is ascribed to traps activated inside the oxide layer above the flakes of the TI and at the interface between the two materials.[18]

A deeper insight on the transport properties is provided by the analysis of the temperature dependence of the device resistance R(T) (see Methods). In the $Bi_2Se_3$ flake, R decreases while increasing T (Fig. 3c) and in the high-temperature range its electrical behavior can be fully explained in terms of an activated transport regime. For T > 40 K, R(T) follows the Arrhenius law $R \propto \exp(\Delta/k_bT)$, where $k_b$ is the Boltzmann constant and Δ is the activation energy which results to be 37 ± 2 meV from a fit to our data. For T < 20 K, R saturates to an approximately constant value (22 MΩ) (Fig. 3d); this suggests that the metallic surface states start to provide a dominant contribution to the overall conductivity.[31] The temperature dependence of the material mobility in the 200 –300 K range follows the power law $T^{-\gamma}$, where $\gamma = 1.8 \pm 0.2$, underlying that the dominant scattering mechanism in the $Bi_2Se_3$ flake, at high temperature, is the electron-acoustic phonon scattering.[32] Conversely, the exfoliated $Bi_2Te_{2.2}Se_{0.8}$ behaves like a highly degenerate semiconductor: R(T) follows indeed the metallic law linear trend $R = R_{rt}\cdot[1+ \alpha(T-T_1)]$, where α is the *positive* thermal coefficient ($\alpha = 1.4\cdot10^{-3}$ K$^{-1}$), $T_1 = 300$ K and $R_{rt}$ is the RT resistance (Fig.3e).

To measure the photoresponse we employed a tunable electronic source operating in the 0.265-0.375 THz range (see Methods). According to the Dyakonov-Shur theory,[3,6] in the case of non-resonant (overdamped) detection, a diffusive model of transport applies (resistive self-mixing regime).[6] The latter predicts a second-order nonlinear response when an oscillating THz field is applied between gate and source. The photovoltage is then:

where the constant η represents the (antenna-dependent) coupling efficiency of the incoming radiation,[6] σ is the overall channel conductivity, $R_{ch}$ is the total S-to-D resistance and $Z_L$ is the (complex) impedance of the readout circuitry. Because of hysteresis effects, photodetection and conductivity were compared for the same sweep. The experimental photovoltages Δu and the resultant responsivities (see Methods) are shown in Fig. 4a and 4b for sample A and sample B, respectively. Equation (1), applied to the up-sweeps plotted in Figs. 3a-b, results in the predicted photovoltage trends shown in the inset of Fig. 4a and on the left vertical axis of Fig. 4d, for sample A and sample B, respectively. The comparison with the corresponding experimental Δu curves, allows identifying the physical mechanisms at the origin of the observed photodetection.

In the case of sample A (Fig. 4a), after an almost flat region, the responsivity $R_v$ (see Methods) decreases while increasing $V_G$, in evident contrast with the overdamped plasma-wave model (inset Fig. 4a).[10] As already demonstrated in graphene FETs,[6] THz detection in an antenna-integrated FET can be, under specific geometrical conditions, mediated by photo-thermoelectric (PTE) effects occurring along the FET channel when the thermal distribution of carriers at the S is “heated” more efficiently by the incoming radiation due to the asymmetrical antenna design (see inset of Fig. 4c). As recently demonstrated,[33] TSS in $Bi_2Se_3$ can enhance PTE effects. The expected PTE voltage ($V_{PTE}$) is proportional to the material Seebeck coefficient ($S_b$) which, in the case of non-degenerate semiconductors with only one carrier type, is related to the charge carrier density ($n$) via the relation:

(2)

Here A is a constant that depends on the dominating scattering mechanism (A = 2 for acoustic phonons), $k_B$ and $h_p$ are the Boltzmann and Planck constants, respectively, $q$ is the carrier charge ($q$<0 for n-type semiconductors) and $\sigma = ne\mu$. In order to infer the dependence of S (and then $V_{PTE}$) from $V_G$, both $\sigma(V_G)$ and $\mu(V_G)$ should be known. At RT and taking into account the small $C_{gc}$ value, combined with the limited gate voltage swing from -10 V to + 10 V, the gate voltage dependence of the material mobility can reasonably be neglected over $\sigma(V_G)$. Under this approximation, $V_{PTE}$ is then estimated as:

(3)

where $\delta T_{el}$ is the electron temperature gradient along the FET channel and the last factor takes into account the loading effects. The gate bias dependence of the ratio $V_{PTE}/\delta T_{el}$ is shown in Fig.4c. A very good qualitative agreement is found between the measured photovoltage Δu (right vertical axis of Fig.4a) and the ratio $V_{PTE}/\delta T_{el}$ (Fig.4c), from which a channel electron temperature gradient of $\delta T_{el} \sim$ 100mK can be inferred. It is worth mentioning that equation (3) does not take into account the possible thermoelectric gate independent shift arising from the ungated regions of the channel. Furthermore, for high positive $V_G$ the Seebeck coefficient approaches zero non-asymptotically (the loading term is ~1 well above threshold): this can be related with the charge density increase above threshold, meaning that the FET material channel cannot be treated here as a non-degenerate semiconductor. In the degenerate case, the $S_b$~ln(1/n) dependence of eq. (2) has to be replaced by $S_b \sim n^{-2/3}$ for which the expected asymptotic behavior would hold.[34]

Conversely, in sample B (Fig. 4b), the experimental $R_v$ curve well agrees with the overdamped plasma-wave predictions (Fig.4d): $R_v$ peaks at high positive $V_G$ although no sign switch is visible at $V_G \sim$ 0V and an offset is present. Both features can be ascribed to a thermoelectric contribution of the ungated regions of the $Bi_2Te_{2.2}Se_{0.8}$ flake, as already shown in graphene THz detectors.[6] For sake of comparison the background signal $\Delta u_{dark}$, measured while blanking the THz beam is shown on the inset of Fig. 4b. To better elucidate the overdamped plasma wave contribution, Fig. 4d (right vertical axis) shows the experimental photovoltage value = $\Delta I \times \sigma^{-1}$ extracted from the difference between the S-D current measured with ($I_{SD,on}$) and without ($I_{SD,off}$) the THz beam and from the measured conductivity σ (Fig. 3b). The comparison with equation (1) (Fig. 4d, left vertical axis) shows an excellent agreement with the theoretical plasma-wave related behavior. Furthermore, by comparing Δu* (Fig. 4d) and Δu (Fig. 4b) we get Δu/Δu* ~ 0.1, as expected if one considers that $\Delta I \times \sigma^{-1}$ is much less influenced by the loading than the photovoltage, since the former is a purely *dc* measurement, not affected by the device capacitive reactance.[7] It is worth mentioning that, in such *dc* configuration, the electrons thermalize more efficiently with the lattice, then reducing significantly the thermoelectric effect.

The achieved overdamped plasma-wave detection in the $Bi_2Te_{2.2}Se_{0.8}$-based FET is mediated by the gated 2D electron gas of the massless Dirac fermions in the TSS.[14,20] This conclusion is supported by the following arguments. From the transport experiments (Figs. 3), we can retrieve the photovoltage value expected in the resistive self-mixing regime (eq. 1) for samples A (inset Fig. 4a) and B (left vertical axis Fig. 4d). The comparison shows that, under almost identical coupling conditions (η value, eq. 1), the expected rectified photovoltage is one-order of magnitude larger in sample A than in sample B. If one assumes that the rectification operated by the self-mixing of over-damped radiation-induced charge density waves occurs in the bulk materials, the rectified photovoltage (eq.1) should be more intense in sample A than in sample B. Conversely, no evidence of plasma-wave detection is visible in sample A. This provides the first experimental indication of plasma-wave detection mediated by TSS in an active TI-based device, and the first optical method to unveil TSS.

Maximum $R_v$ of 3.0 V/W and 0.21 V/W have been reached for sample A and B, respectively. Under the assumption that the detector noise figure is dominated by the thermal Johnson-Nyquist contribution $N_{th} = (4R_{ch}k_BT)^{1/2}$, we can infer the noise-equivalent power (NEP) i.e. the lowest detectable power in a 1 Hz bandwidth that can be calculated as $N_{th}/R_v$.[4,6] The noise was then estimated by measuring the resistance of the device while the THz beam was impinging on it. Figures 5a-b show the extrapolated NEP values as a function of $V_G$. In both cases, a minimum NEP of ≈ 10 nW/√Hz has been found. Despite the lower responsivity of sample B, its 300 times larger conductivity is beneficial to improve the signal-to-noise ratio, thus making such detectors already exploitable for practical applications. To provide a concrete exploitation in a quality control application, we have then employed sample B in a transmission imaging experiment. The 0.33 THz beam was focused on the target object and the transmitted power was detected in photovoltage configuration while keeping $V_G$ = 0 V (all electrodes were unbiased). The 400 × 700 pixel image of a target object was acquired with a time constant of 20 ms. The THz scan is reported in Fig. 5c: the glue jar (Fig. 5d) is not completely full and the quantity of liquid, as well as the air gap in between, is clearly resolved by our THz detector with a remarkable high signal-to-noise ratio SNR ~ 1000, as extrapolated from the comparison between the background signal $\Delta u_{dark}$ and the recorded photovoltage (Fig.4b). The achieved results represent a crucial milestone in the road map for TI-based science and technology.

## Methods

### Growth, material characterization and device fabrication

The polycrystalline samples, synthesized from high-purity starting elements, were placed in a conical-bottom quartz ampoule, which was sealed under vacuum conditions. Before the growth process, the ampoule was held in the "hot" zone (1050 K) of the two-zone tube furnace for 12 h for a complete melting of the composition. Then, the charged ampoule was moved from the "hot" zone to the "cold" zone (~ 900 K) with a rate of 1.0 mm/h. The as-grown crystals, consisting of one large single-crystalline block, were then mechanically exfoliated on a 350 μm thick silicon wafer with a 300 nm thin insulating $SiO_2$ top-layer.

The AFM mapping has been performed by employing a Bruker system (IconAFM) with thickness resolution < 1 nm and lateral resolution ~20 nm. The Raman spectra of the exfoliated flakes were collected by exciting the samples with the 532 nm line of a Nd-Yag pumping laser with an optical power density of 0.4 mW/μm$^2$, exploiting a spatial lateral resolution of 2 μm. Due to the very poor thermal conductivity of both crystals, the Raman measurement is likely to damage the flakes even at low incident powers. Therefore, different flakes have been used for crystal characterization and device fabrication, respectively. The mechanical exfoliation of both $Bi_2Se_3$ and $Bi_2Te_{2.2}Se_{0.8}$ crystals resulted in flakes of lateral size in the 1-8 μm range and thicknesses (*h*) in the 10-100 nm range. The Raman spectra show good crystalline quality for both materials. In the case of $Bi_2Se_3$ (Fig.1-e) the $E_g^1$, $A_{1g}^1$, $E_g^2$, $A_{1g}^2$ vibrational modes are clearly visible at wavenumbers of 36 cm$^{-1}$, 72 cm$^{-1}$, 130 cm$^{-1}$ and 172 cm$^{-1}$, respectively; alternatively, for $Bi_2Te_{2.2}Se_{0.8}$ (Fig.1-f) the Raman spectrum shows the same set of peaks, but at wavenumbers of 36 cm$^{-1}$, 62 cm$^{-1}$, 101 cm$^{-1}$ and 135 cm$^{-1}$.

Thin flakes (*h* ~ 10 nm, corresponding to ~ 7 quintuple layers) were individually contacted with Cr/Au (sample A) or Pd (sample B) to define the S and D electrodes via aligned electron beam lithography (EBL). Although the use of an adhesion layer (like Cr or Ti) could induce deterioration and a related rapid increase in contact resistance we did not found any noticeable differences in device performances in time. The S electrode was then patterned in the shape of a half 100° bow-tie antenna, introducing the required asymmetry for THz detection. A 60 nm thick layer of $SiO_2$ was sputtered over the samples to form the top-gate oxide; we found that this technique provides good uniformity if the oxide is deposited at pressures below $10^{-2}$ mbar. The G electrode, designed with the same geometry of the S electrode, was aligned with the center of the channel via EBL and defined via thermal evaporation of 80 nm layer of Cr/Au. Sample A is equipped with a bow-tie antenna with radius $L$ = 250 μm, while the bow radius of sample B is L = 500 μm, i.e. corresponding to the two strongest resonances of the designed antenna (2L = λ/2 and 2L = λ) at 0.3 THz. For sample A, the channel length is $L_{sd}$ = 500 nm and the gate length is $L_g$ = 220 nm while for sample B, $L_{sd}$ = 1.2 μm and $L_g$ = 400 nm. These dimensions lead to different values of the geometrical gate-to-channel capacitance ($C_{gc}$): the simulated $C_{gc}$ values are $C_{gc} \approx 100$ aF and $C_{gc} \approx 240$ aF for sample A and sample B, respectively.

**Transport and optical characterization**

The temperature dependence of the resistance of sample A and B has been measured by using a liquid helium cryostat equipped with a calibrated silicon diode as thermometer.

The optical characterization has been performed by focusing the THz frequency beam on a spot of 4 mm diameter at the detector surface by means of a set of *f*/1 off axis parabolic mirrors and mechanically chopped at 619 Hz. The THz radiation was linearly polarized along the axis of the bow-tie antennas in order to maximize the photoresponse signal.[4] The power of the source ($P_t$), calibrated as a function of output frequency with a power meter, ranges between 200 μW and 450 μW. In our experiment, a photovoltage-mode (PV) configuration was employed: the S electrode was grounded, $V_G$ was set with a Keithley *dc* generator and Δu was measured at the D electrode with a lock-in amplifier (Fig. 2e). A low noise voltage pre-amplifier (input impedance = 10 MΩ) with a pass-band filter between 300 Hz and 1 kHz was used in the experiments with gain factor $G_n$ = 250 for sample A and $G_n$ = 1000 for sample B. From this voltage measurement Δu can be calculated using the equation:[4]

where *LIA* is the voltage read by the lock-in, and the factor π√2/2 is a normalization that takes into account that the lock-in measures the rms of the fundamental sine wave Fourier component of the square wave produced by the chopper.

We preventively measure Δu as a function of frequency by employing the tunable source to identify, in both cases, the antenna resonances: we found that $\nu_A$ = 292.7 GHz ($P_t$ = 370 μW) (sample A) and $\nu_B$ = 332.6 GHz ($P_t$ = 330 μW) (sample B) correspond to the highest intensity response peaks that we selected as operating conditions. The responsivity ($R_v$) was then determined using the relation $R_v = (\Delta u \cdot S_t)/(P_t \cdot S_a)$, where $S_t$ is the beam spot area and $S_a$ is the active area of the detector; in the case of sample A, $S_a$ is set equal to the diffraction limited area ($S_\lambda = \lambda^2/4$), being the antenna surface smaller than $S_\lambda$.[4]

To assess the THz induced photocurrent we exploited a photoconductive scheme (Fig. 2e): a *dc* bias is applied to the source electrode ($V_{SD}$) and the current is read at the D electrode with an amperometer while shining the THz radiation on the detector ($I_{SD,on}$) or keeping the beam blanked ($I_{SD,off}$). The photocurrent is then estimated as the difference between the currents measured in the two configurations: $I_P = I_{SD,on} - I_{SD,off}$.

**Acknowledgments**

We acknowledge funding from: the Italian Ministry of Education, University, and Research (MIUR) through the program "FIRB - Futuro in Ricerca 2010 RBFR10LULP "Fundamental research on Terahertz photonic devices", the European Union through the MPNS COST Action"MP1204 TERA-MIR Radiation: Materials, Generation, Detection and Applications", the National Science Poland Centre (DEC-2013/10/M/ST3/00705), from the Basque Country Government, Departamento de Educación, Universidades e Investigación (Grant No. IT-756-13), the Spanish Ministerio de Ciencia e Innovación (Grant No. FIS2010-19609-C02-01), the Saint Petersburg State University (Project No. 11.50.202.2015), the Science Development Foundation under the President of the Republic of Azerbaijan, Grant No. EIF-2011-1(3)-82/69/4-M-50. W.K and M.S.V acknowledge F. Teppe for fruitful discussions.

## Figure captions

**Figure 1**: **Flake identification after mechanical exfoliation**. (a,b) Scanning electron microscopy (SEM) images of exfoliated flakes of $Bi_2Se_3$, and $Bi_2Te_{2.2}Se_{0.8}$, respectively; brighter flakes correspond

to thicker flakes. (c,d) Atomic force microscopy (AFM) images: the thickness of exfoliated flakes spans from 10 to 100 nm. (d) AFM scan of a single quintuple layer (thickness ~1.4 nm) (e,f) Micro-Raman spectra measured using the 532 nm line of a Nd-Yag pumping laser. For $Bi_2Se_3$ (e) the Raman spectrum shows peaks at 36 $cm^{-1}$, 72 $cm^{-1}$, 130 $cm^{-1}$ and 172 $cm^{-1}$, corresponding to the $E_g^1$, $A_{1g}^1$, $E_g^2$, $A_{1g}^2$ vibrational modes, respectively. (f) The Raman spectrum of $Bi_2Te_{2.2}Se_{0.8}$ shows peaks at 36 $cm^{-1}$ ($E_g^1$), 62 $cm^{-1}$ ($A_{1g}^1$), 101 $cm^{-1}$ ($E_g^2$), 135 $cm^{-1}$ ($A_{1g}^2$).

**Figure 2**: **Device Fabrication.** (a,b) False colors SEM images of the top-gated FETs in sample A (a) and sample B (b), respectively. (c,d) False colors SEM images of the patterned bow-tie antennas in sample A (2L = 0.5 mm) and sample B (2L = 1 mm), respectively. (e) Schematics of the THz detection principle, photovoltage and photocurrent modes are depicted in black and purple, respectively.

**Figure 3**: **Transport characterization.** (a) Sample A: Transconductance $I_{SD}(V_G)$ measured at RT in a $Bi_2Se_3$ based FET, sweeping the gate bias ($V_G$) in the range (-10 V, +10 V) and keeping the source-drain voltage $V_{SD}$ = 1mV. (b) Sample B: $I_{SD}(V_G)$ measured at RT in a $Bi_2Te_{2.2}Se_{0.8}$ based FET, sweeping $V_G$ in the range (-7 V, +7 V) and keeping the source-drain voltage $V_{SD}$ = 1mV. (c) Sample A: temperature dependence of the resistance (R). Insets: (left panel) logarithmic plot of the conductance, above 40 K the resistance follows the Arrhenius law (activated transport regime with $\Delta$ = 37 ± 5 meV, fitted by the red line), for T < 20 K, R exhibits a metallic trend (blue line); (right panel) field effect mobility measured between 200 and 300 K as a function of the temperature. The fit to the data gives $\mu(T) \propto T^{-1.8}$. (d) Sample B: temperature dependence of the resistance. R(T) follows a linear trend typical of degenerate semiconductors with $\alpha = 1.4 \cdot 10^{-3}\ K^{-1}$ (the dashed line is a linear fit to the data).

**Figure 4: Optical characterization**. (a) Sample A: Experimental responsivity (left vertical axis) and photovoltage (right vertical axis) measured at room-temperature at a fixed frequency $\nu$ = 292.7 GHz while sweeping the gate bias $V_G$ in the range (-10 V, +10 V). Inset: theoretical photovoltage ($\Delta u_{theory}$) predicted by the Diakonov-Shur overdamped plasma-wave theory divided by the antenna coupling constant ($\eta$). (b) Sample B: Experimental responsivity (left vertical axis) and photovoltage (right vertical axis) measured at RT with a 332.6 GHz radiation while sweeping the gate bias $V_G$ in the range (-7 V, +7 V). Inset: Background signal ($\Delta u_{dark}$) measured while keeping the 332.6 GHz beam blanked. (c) Calculated PTE voltage ($V_{PTE}$) for sample A. Inset: schematics of the main physical mechanism. (d) Left vertical axis: theoretical photovoltage ($\Delta u_{theory}$) predicted by the Diakonov-Shur overdamped plasma-wave theory divided by the antenna coupling constant ($\eta$). Right vertical axis: difference between the source-drain currents measured while focusing the THz beam on the sample ($I_{SD,on}$) and without it ($I_{SD,off}$), multiplied by the channel resistance. Inset: transconductance curve of sample B collected while shining the 332.6 GHz radiation on the detector and while the THz beam is blanked (red and purple curves in inset) and while keeping $V_{SD}$ = 0.6 mV.

**Figure 5**. **Fast, large-area, RT, THz imaging**. (a) Sample A: noise equivalent power (NEP) plotted as a function of the gate bias $V_G$ for a 292.7 GHz impinging radiation frequency. (b) Sample B: noise equivalent power (NEP) plotted as a function of $V_G$ obtained for a 332.6 GHz impinging radiation frequency. (c) RT, large area THz imaging obtained while impinging the 332.6 GHz radiation on sample B, mounted on a *XY* stage, with an acquisition time of 20 ms/pixel. For visible light illumination the contents cannot be seen, either by naked eye or by the CCD camera used to take the picture. The detection of THz transmitted radiation gives information about the jar content. (d) Photograph of the glue jar.

**Figure 1**

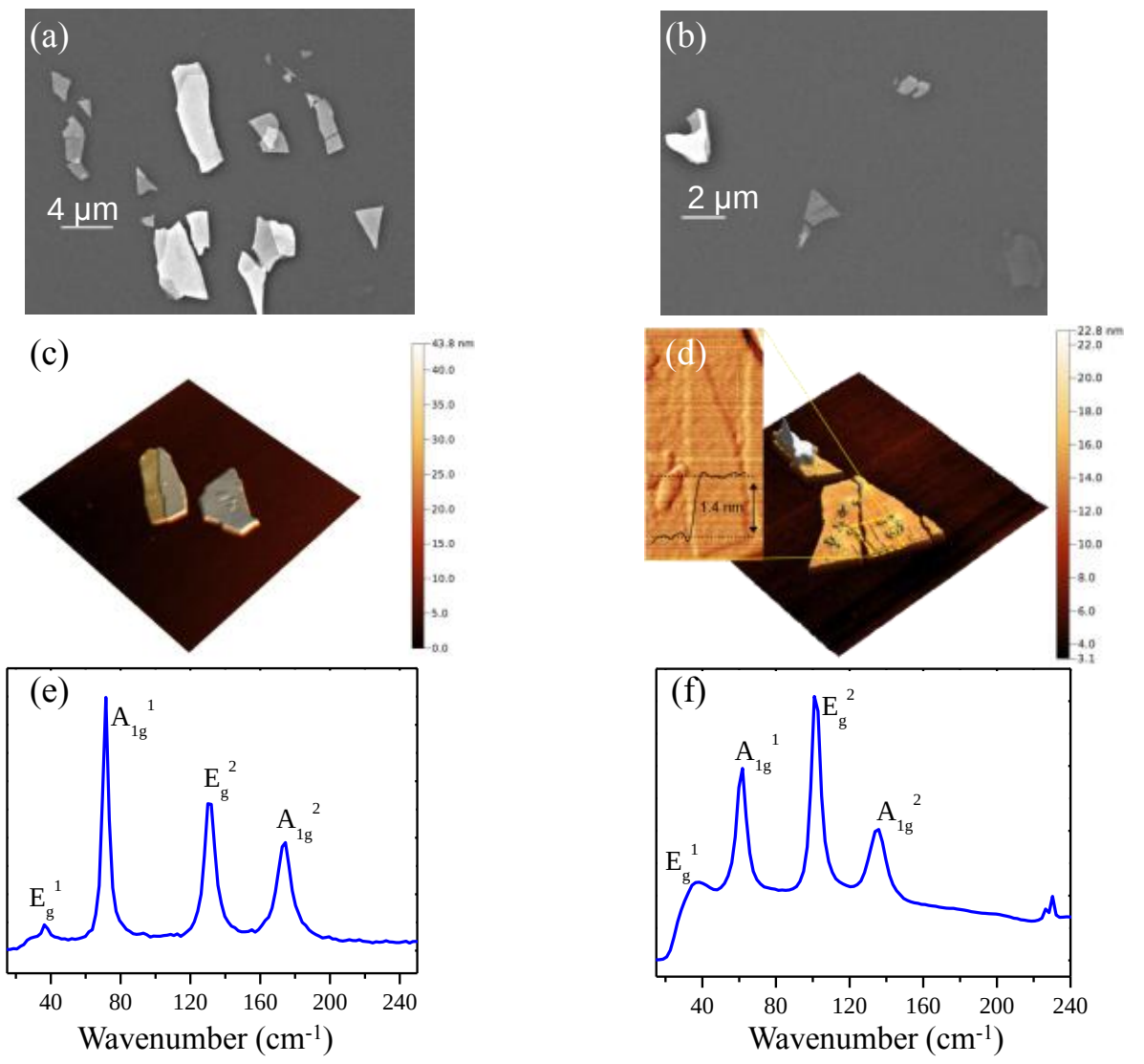

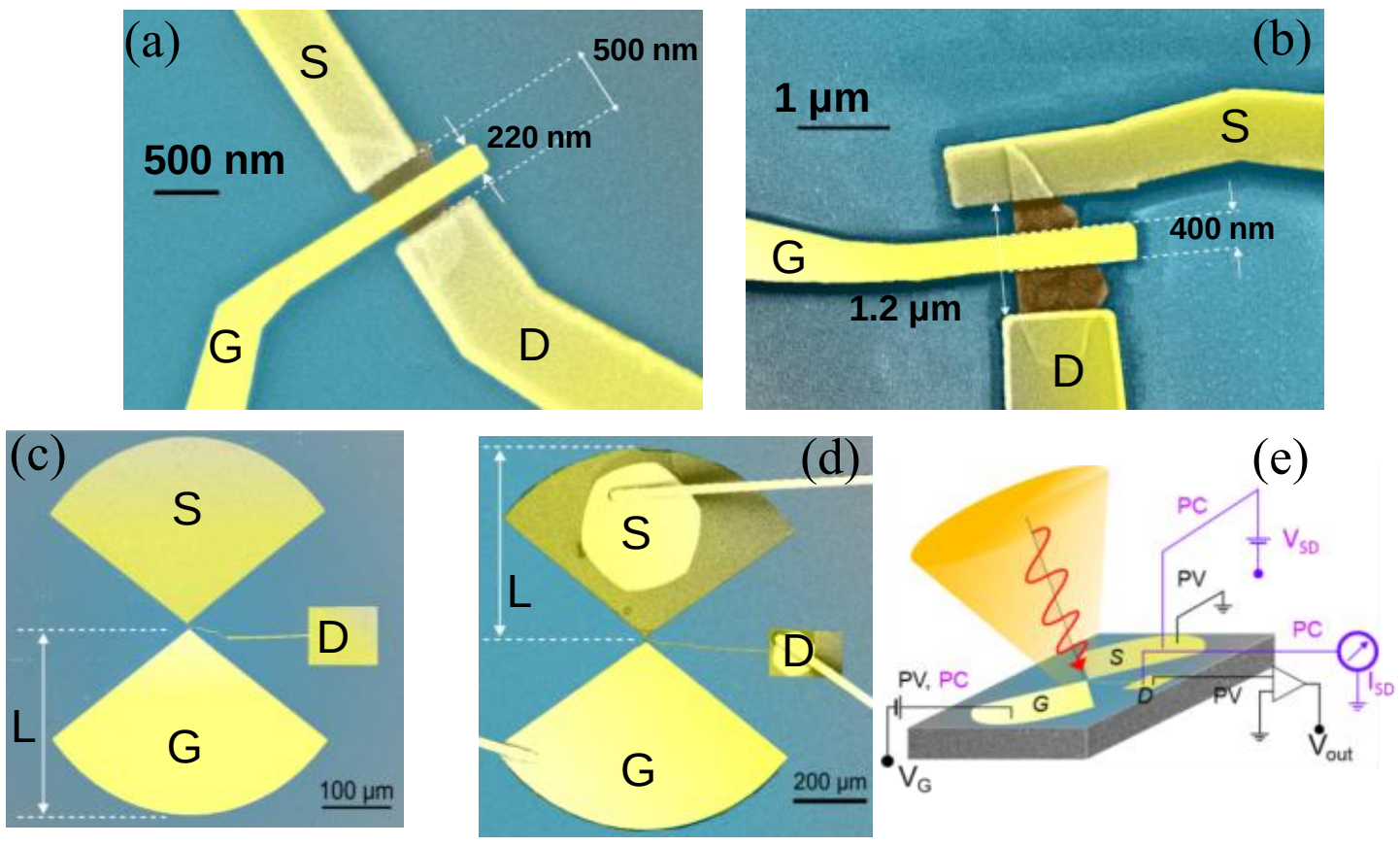

**Figure 3**

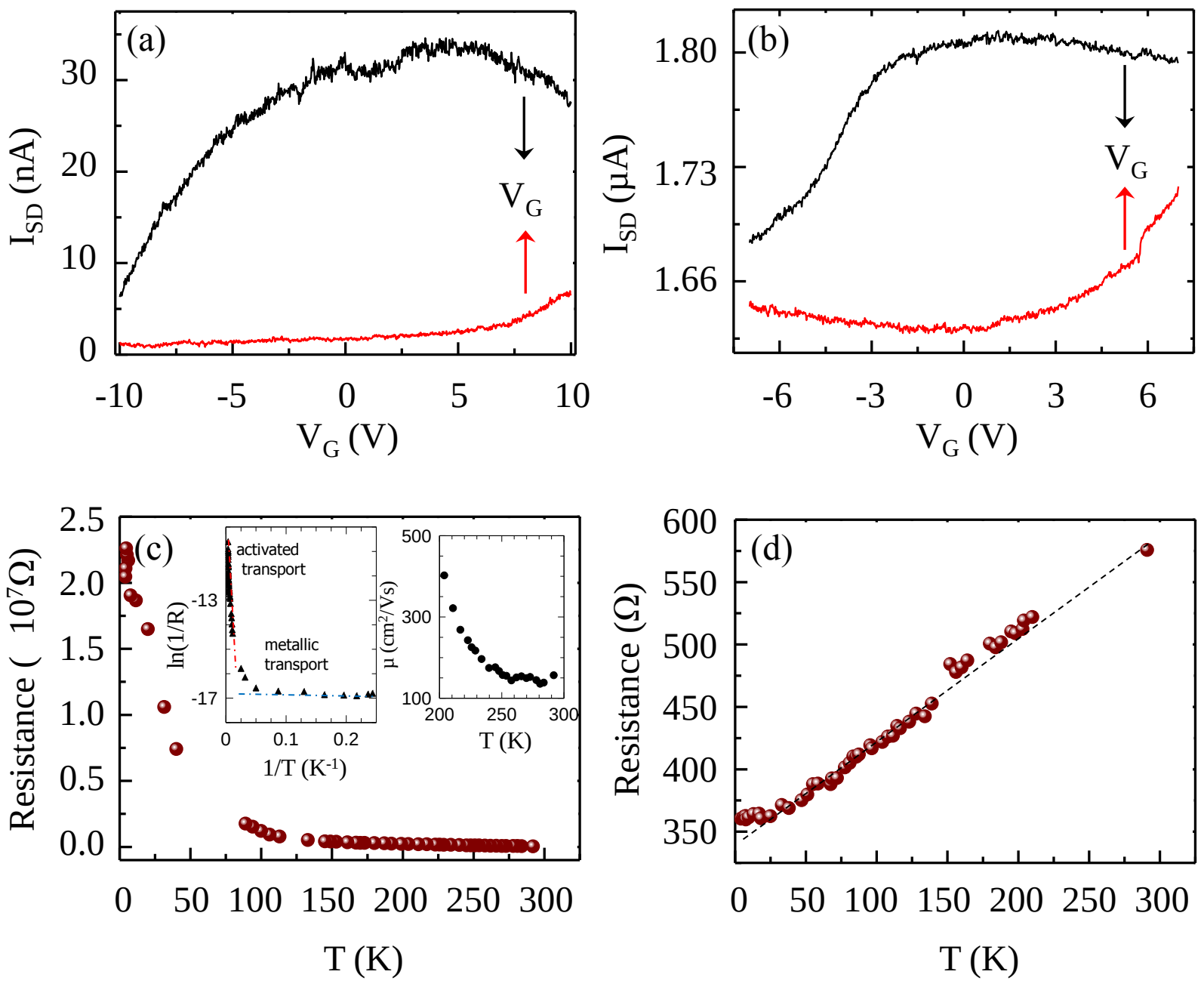

**Figure 4**

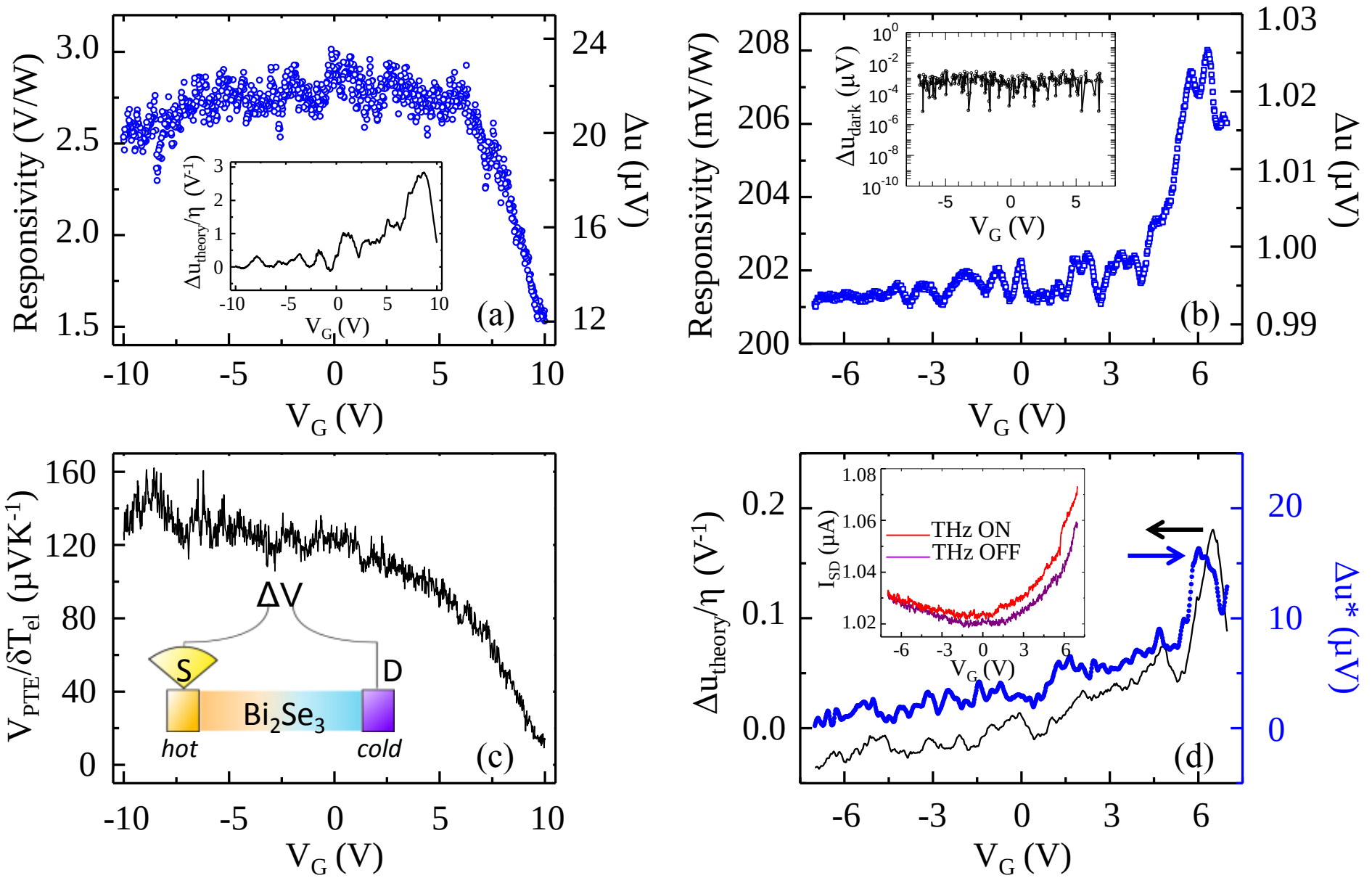

**Figure 5**

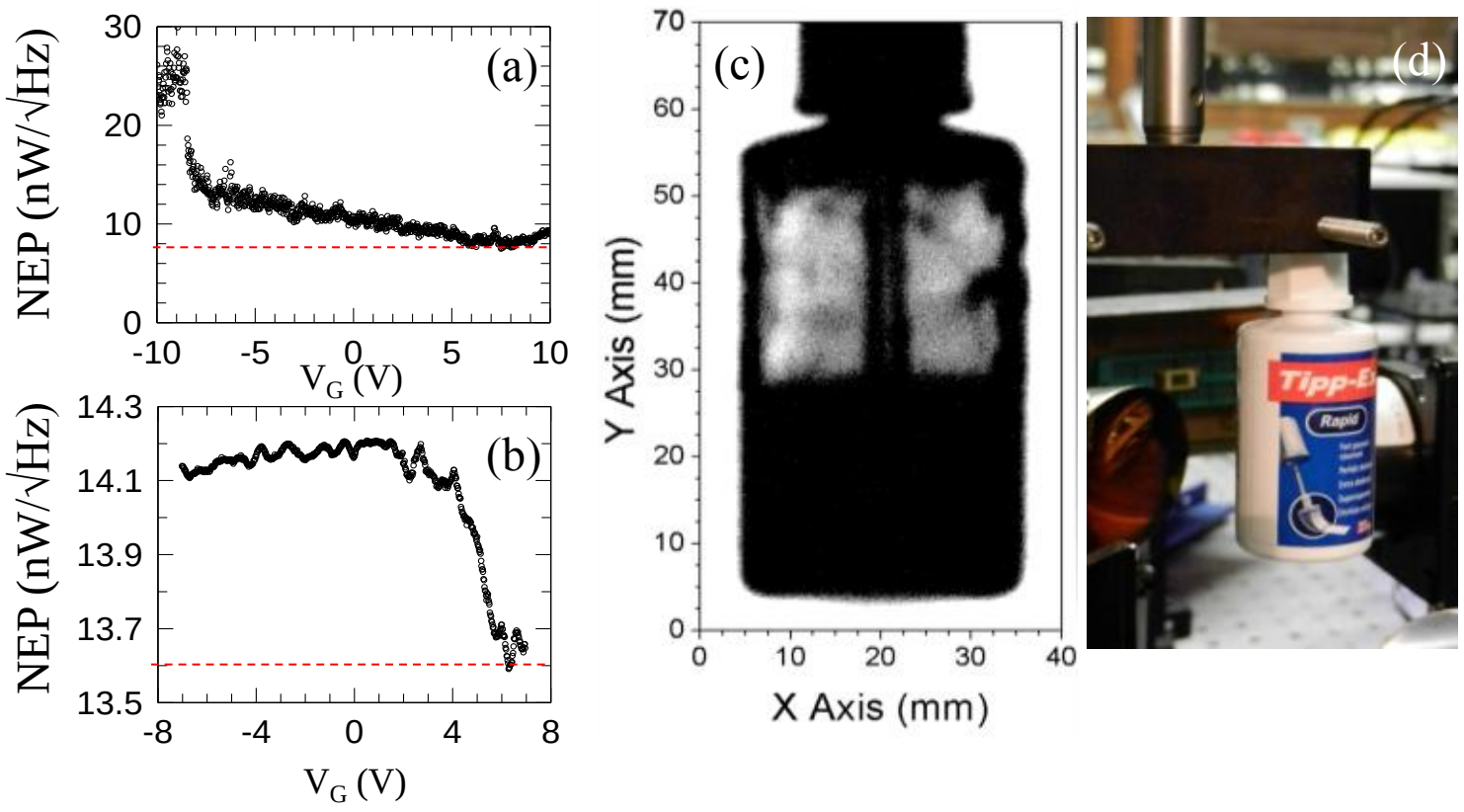